\documentclass[acmlarge,screen]{acmart}
\usepackage{import}
\usepackage{multirow, multicol}
\usepackage{array}
\usepackage{subfigure}
\usepackage{footnote}
\AtBeginDocument{%
  \providecommand\BibTeX{{%
    \normalfont B\kern-0.5em{\scshape i\kern-0.25em b}\kern-0.8em\TeX}}}

\acmConference[Workshop on Generative AI and HCI]{}{CHI '22}{May 10}

\begin{document}

\title{Subjective Evaluation of Deep Learning Models for Symbolic Music Composition}

\author{Carlos Hernandez-Olivan}
\authornote{carloshero@unizar.es \\ \url{https://carlosholivan.github.io}}
\email{carloshero@unizar.es}
\orcid{0000-0002-0235-2267}
\author{Jorge Abadias Puyuelo}
\email{759704@unizar.es}
\author{Jose R. Beltran}
\email{jrbelbla@unizar.es}
\orcid{0000-0002-7500-4650}
\affiliation{%
  \institution{Department of Engineering and Communications, University of Zaragoza}
  \streetaddress{María de Luna, 3}
  \city{Zaragoza}
  \state{Spain}
  \country{Spain}
  \postcode{50018}
}

\renewcommand{\shortauthors}{Hernandez-Olivan, et al.}

\begin{abstract}
  Deep learning models are typically evaluated to measure and compare their performance on a given task. The metrics that are commonly used to evaluate these models are standard metrics that are used for different tasks. In the field of music composition or generation, the standard metrics used in other fields have no clear meaning in terms of music theory. In this paper, we propose a subjective method to evaluate AI-based music composition systems by asking questions related to basic music principles to different levels of users based on their musical experience and knowledge. We use this method to compare state-of-the-art models for music composition with deep learning. We give the results of this evaluation method and we compare the responses of each user level for each evaluated model.
\end{abstract}

\ccsdesc[500]{Computing methodologies}
\ccsdesc[500]{Computing methodologies~Artificial intelligence}
\ccsdesc[500]{Applied computing~Sound and music computing}

\keywords{music composition, music information retrieval, artificial intelligence, machine learning, deep learning, neural networks, subjective evaluation, music generation evaluation}

\maketitle

\section{Introduction} \label{sec:creativity}

Music composition or generation is a subfield of the Music Information Retrieval (MIR) that aims to create new music in the symbolic or in the audio domains. Music composition can be defined as the act of conceiving a piece of music or the art of creating music. As the definition claims, the process of making a new music piece is a creative process. Music Metacreation (MuMe) is the subfield of computational creativity that addresses music-related creative tasks \cite{pasquier2017introduction}. The process that a human follows to create a piece of music depends on a wide variety of decisions that the composer take based on his or her music knowledge, experience or subjective taste \cite{hernandezolivan2021music}. The performance of generative models is often preferable to be measured by subjective evaluation rather than objective evaluation \cite{yang2020evaluation}. Music composition is evaluated with listening tests which is not an easy task. There are risks that may lead into incoherences in the evaluation systems such as the confusion between the question about the aesthetic of the music piece or the question about the entity who composed the piece, AI or human. Some listening tests ignore the music knowledge of the users, and others only rely on the preference of one model over another \cite{yang2020evaluation}.

\section{Method and Results} \label{sec:eval}

AI-based music composition is normally evaluated with metrics such as the accuracy or the perplexity. DL models are also compared with their loss function. The loss function can give a general idea of how the model generalizes in a particular dataset, but there is a lack of standard method to evaluate and properly compare AI-based music composition models due to the difficulties of measuring music. Accuracy metrics cannot explain or measure all the music intricacies that depend on the music basic principles. We can distinguish between subjective and objective evaluation methods. In one hand, there have been done listening tests \cite{musicvae}, \cite{deepbach} that are too general to evaluate the music that a model generates. In the other hand, there have been proposed some objective metrics \cite{survey}, \cite{yang2020evaluation}. Symbolic music basic principles are: melody, harmony, form or structure and texture. There are other music principles that are not in the scope of this work because they are not sheet or symbolic music principles but they are more related to the performance such as timbre, dynamics or effects. In Table \ref{tab:survey} we tag the survey questions to these principles: melody (M), harmony (H) and rhythm (R). Although creativity tests do not ensure a general indicative of creativity \cite{amabile2018creativity}, the online survey with a Turing test \cite{turing2009computing}, \cite{ariza2009interrogator} or questions are the most-common technique to evaluate the results of a music AI-based generation system. Our work presents a subjective way of evaluation AI-based music composition systems.

\subsection{Surveys}
Normally, surveys fail because the experience level of the users is not taken into account \cite{yang2020evaluation}. To avoid that, we divided our surveys by levels of music knowledge of the users with different questions regarding each user-level music experience. We based our survey's questions in MuseGAN user study \cite{musegan}. Questions are rated in a 5-point Likert scale. We also ask users to distinghuish between AI or human composition as we included human-made dataset samples in the surveys. In Table \ref{tab:survey} we propose a template of questions divided in three different levels depending on the music knowledge of the users. Each survey contains one piece generated with each of the models that are evaluated in this work and one piece taken directly from the datasets with which the models have been trained with. This means that each user listens to 8 samples of approximately 30 seconds each to avoid listening fatigue.
We propose 3 user levels:

\begin{itemize}
    \item Basic user. People with no music knowledge or amateur musicians.
    \item Middle user. People with some music knowledge such as conservatory students.
    \item Pro user. Professional musicians with a strong background in music theory.
\end{itemize}

\newcommand{\specialcell}[2][c]{%
  \begin{tabular}[#1]{@{}c@{}}#2\end{tabular}}
  
\begin{table*}[h]
    \centering
    \caption{Online survey questions per user level. M: melody, R: rhythm and H: harmony related questions.}
    \begin{tabular}{
    >{\centering\arraybackslash}p{2cm}  
    m{10.5cm} 
    >{\centering\arraybackslash}p{2.5cm} 
    }
    \toprule
        User & Question & Score range\\
        \midrule
        \multirow{4}{*}{Basic} & \multirow{1}{*}{Q1. (\textbf{M, H, R}) I feel I have heard similar music before.} & \multirow{1}{*}{1-5}\\
        & \multirow{1}{*}{Q2. (\textbf{M, H, R}) The piece conveys something to me.} & \multirow{1}{*}{1-5}\\
        & \multirow{1}{*}{Is this sample composed by a human or is it AI-generated?} & \multirow{1}{*}{AI/HC/NS}\\
        & \multirow{1}{*}{Musical composition overall rating.} & \multirow{1}{*}{1-5}\\

        \midrule
        
        \multirow{6}{*}{Middle} & \multirow{1}{*}{Q1. (\textbf{M, H, R}) I feel I have heard similar music before.} & \multirow{1}{*}{1-5}\\
        & \multirow{1}{*}{Q2. (\textbf{H}) The harmony is pleasant.} & \multirow{1}{*}{1-5}\\
        & \multirow{1}{*}{Q3. (\textbf{R}) I find some irregularities in the rhythm.} & \multirow{1}{*}{1-5}\\
        & \multirow{1}{*}{Q4. (\textbf{M, H, R}) Could you identify the music style or genre of the piece?} & \multirow{1}{*}{Yes/No/Not Sure}\\
        & \multirow{1}{*}{Is this sample composed by a human or is it AI-generated?} & \multirow{1}{*}{AI/HC/NS}\\
        & \multirow{1}{*}{Musical composition overall rating.} & \multirow{1}{*}{1-5}\\
        
        \midrule
        
        \multirow{9}{*}{Pro} & \multirow{1}{*}{Q1. (\textbf{R}) The tempo is coherent with the melody or music genre.} & \multirow{1}{*}{1-5}\\
        & \multirow{1}{*}{Q2. (\textbf{M, R}) There is an identifiable musical motif.} & \multirow{1}{*}{1-5}\\
        & \multirow{1}{*}{Q3. (\textbf{R}) I find some irregularities in the rhythm.} & \multirow{1}{*}{1-5}\\
        & \multirow{1}{*}{Q4. (\textbf{H}) There is an identifiable harmonic progression.} & \multirow{1}{*}{1-5}\\
        & \multirow{1}{*}{Q5. (\textbf{M, H, R}) Can you identify the period or music style of the piece?} & \multirow{1}{*}{Yes/No/Not Sure}\\
        & \multirow{1}{*}{Q6. (\textbf{H}) Would you say it's tonal music?} & \multirow{1}{*}{Yes/No/Not Sure}\\
        & \multirow{1}{*}{Is this sample this sample composed by a human or is it AI-generated?} & \multirow{1}{*}{AI/HC/NS}\\
        & \multirow{1}{*}{Musical composition overall rating.} & \multirow{1}{*}{1-5}\\

         \bottomrule
    \end{tabular}
    \label{tab:survey}
\end{table*}

\subsection{Evaluated Models}
The state-of-the-art evaluated models in this work are: MusicVAE (MVAE) \cite{musicvae}, Music Transformer \cite{music_transformer}, DeepBach \cite{deepbach} and the Multi-Track Music Machine (MMM) \cite{ens2020mmm}. Each one is build with a different neural network architecture: VAEs \cite{vae}, Transformers \cite{attention} and recurrent neural networks \cite{errorpropnonote} with the Gibbs sampling.
MusicVAE and MMM are trained with the Lakh MIDI dataset (LMD) \cite{raffel2016learning}, the Music Transformer is trained with the MAESTRO dataset \cite{maestro} and DeepBach uses the JSB Chorales dataset (JSBC) \cite{jsbchorales} in its training. These models generate music for different ensembles: MusicVAE for trios, MMM for a custom ensemble, Music Transformer for piano and DeepBach for Bach chorales. We choose the "from scratch" generation in the models that can also be conditioned.

\subsection{Results}
\begin{figure}[ht]
    \begin{minipage}[b]{0.34\textwidth}
        \centering
        \includegraphics[width=\columnwidth]{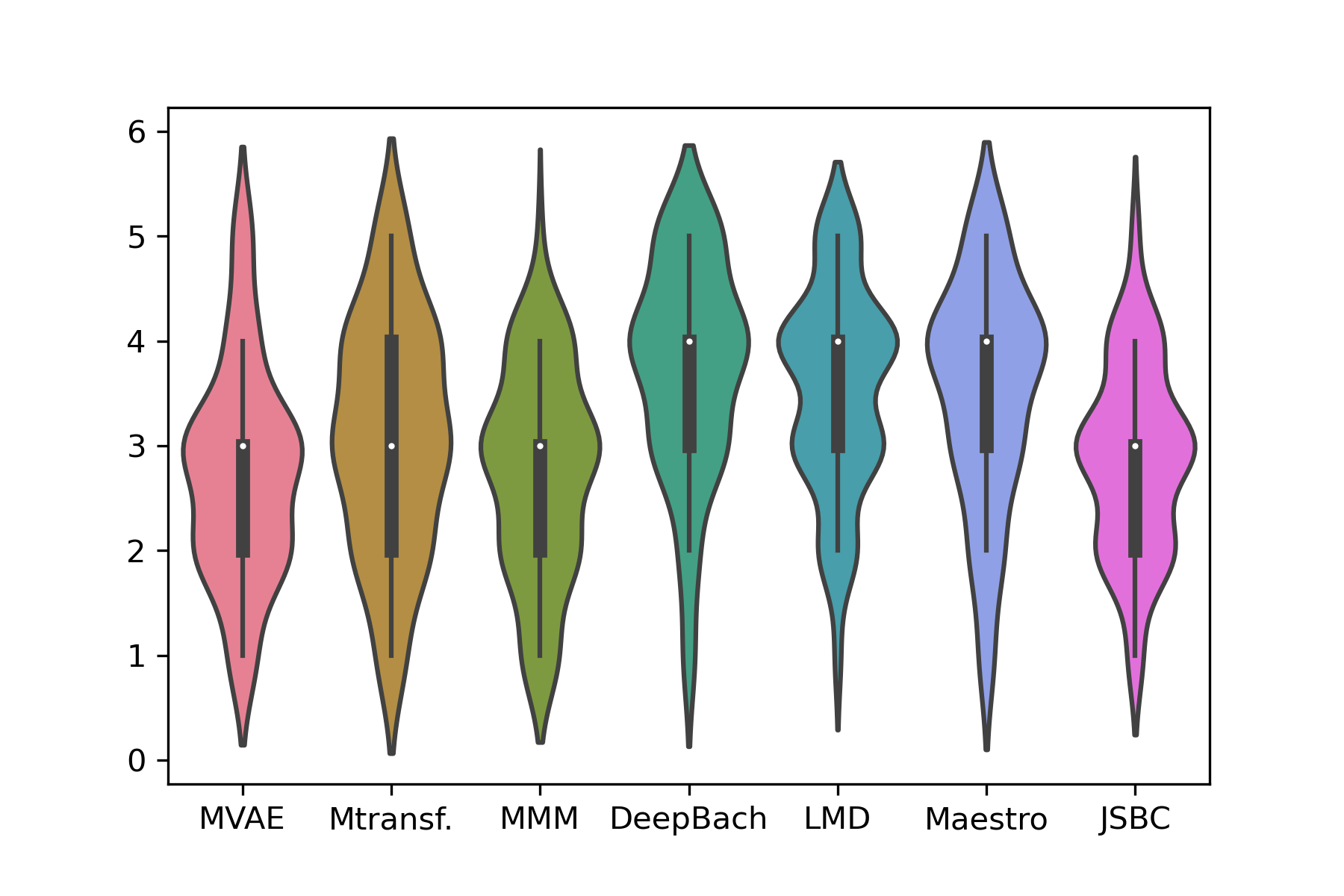}\\
        \caption*{a) Beginner user}
    \end{minipage}%
    \begin{minipage}[b]{0.34\textwidth}
        \centering
        \includegraphics[width=\columnwidth]{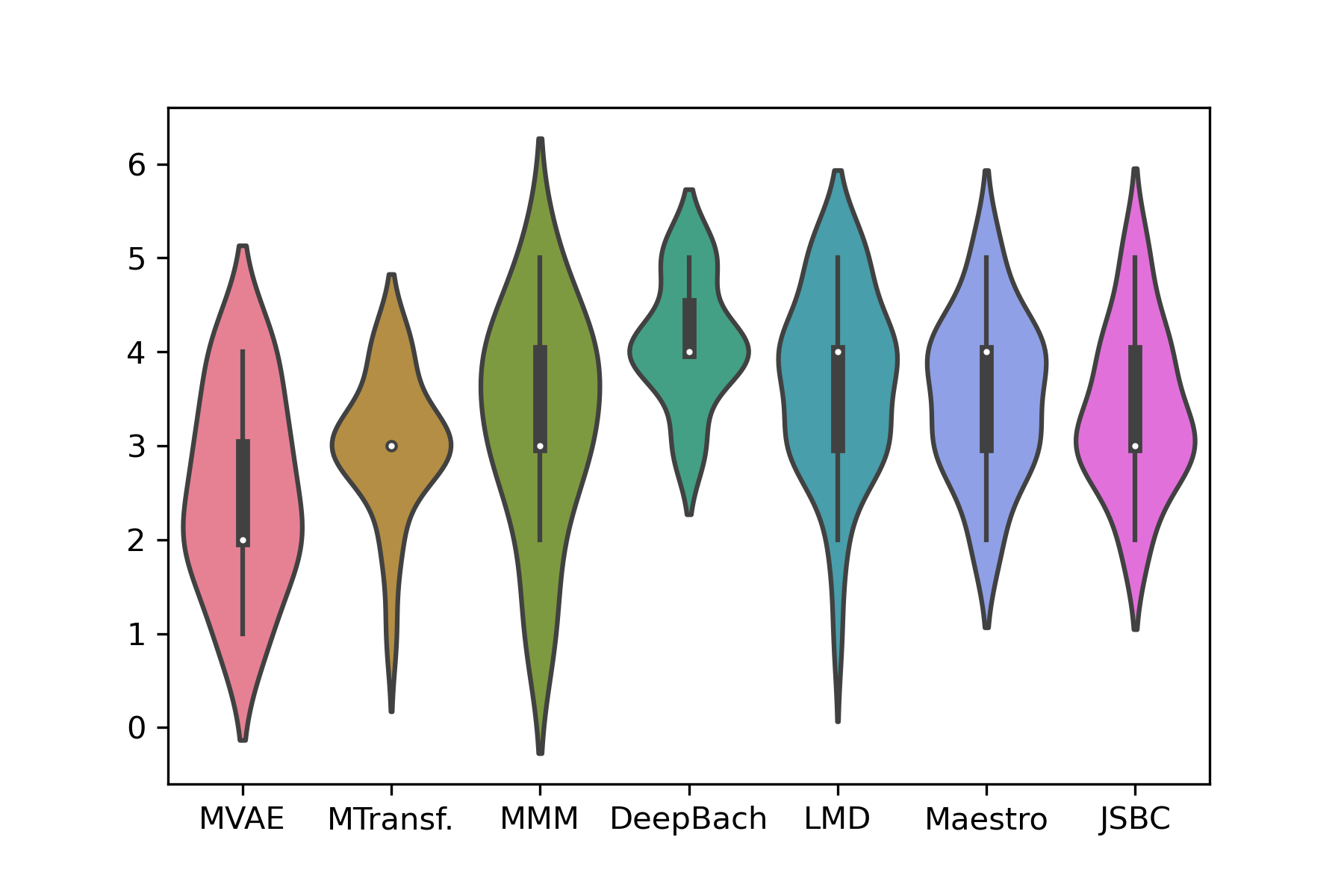}\\
        \caption*{b) Intermediate user}
    \end{minipage}%
    \begin{minipage}[b]{0.34\textwidth}
        \centering
        \includegraphics[width=\columnwidth]{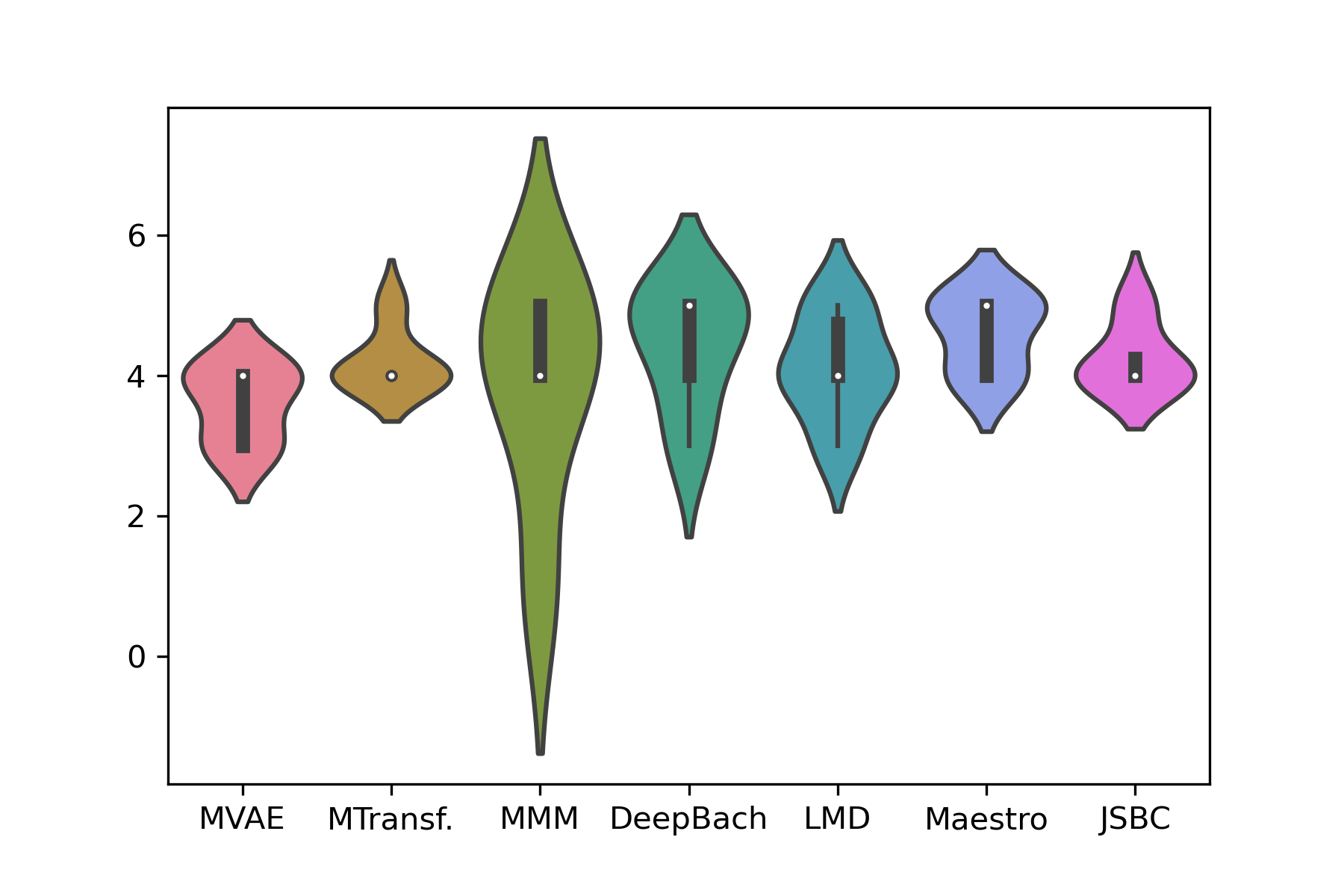}\\
        \caption*{c) Pro user}
    \end{minipage}%
\caption{Overall rating results for different user levels. AI refers to Artificial Intelligence, HC to human composition and NS to not sure.}
\label{fig:overall}
\end{figure}

\begin{figure}
    \begin{minipage}[b]{0.34\textwidth}
        \centering
        \includegraphics[width=\columnwidth]{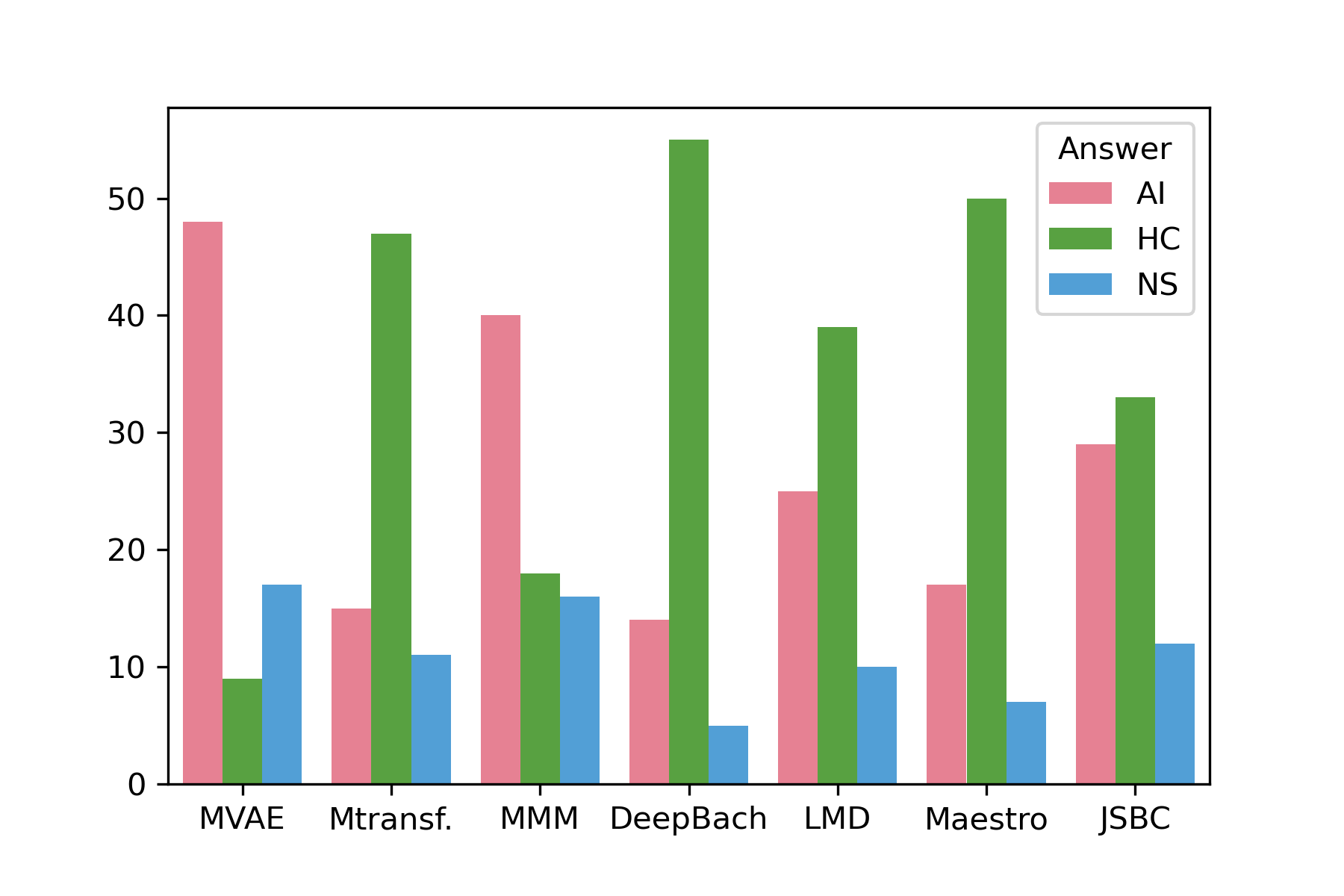}\\
        \caption*{a) Beginner user}
    \end{minipage}%
    \begin{minipage}[b]{0.34\textwidth}
        \centering
        \includegraphics[width=\columnwidth]{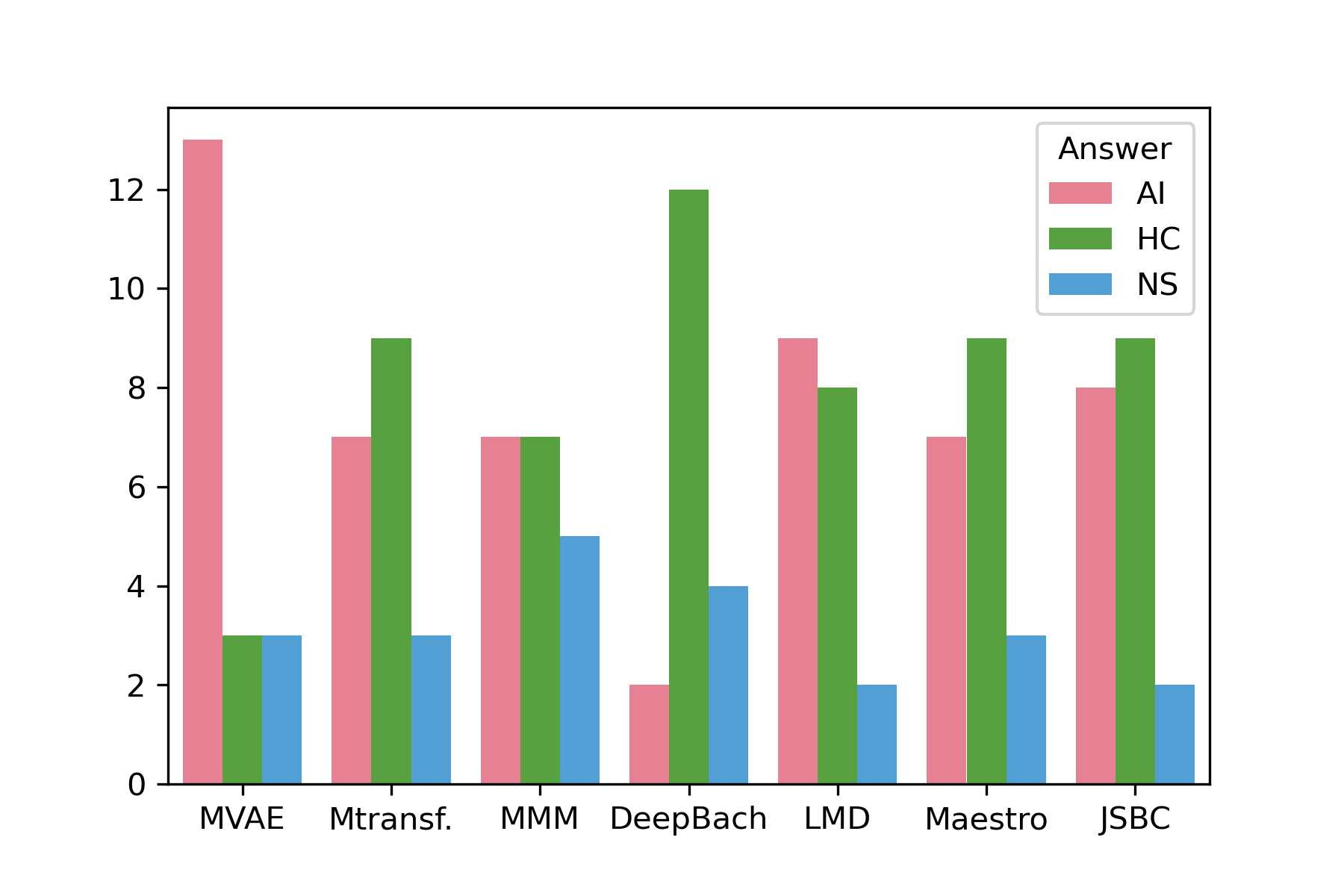}\\
        \caption*{b) Intermediate user}
    \end{minipage}%
    \begin{minipage}[b]{0.34\textwidth}
        \centering
        \includegraphics[width=\columnwidth]{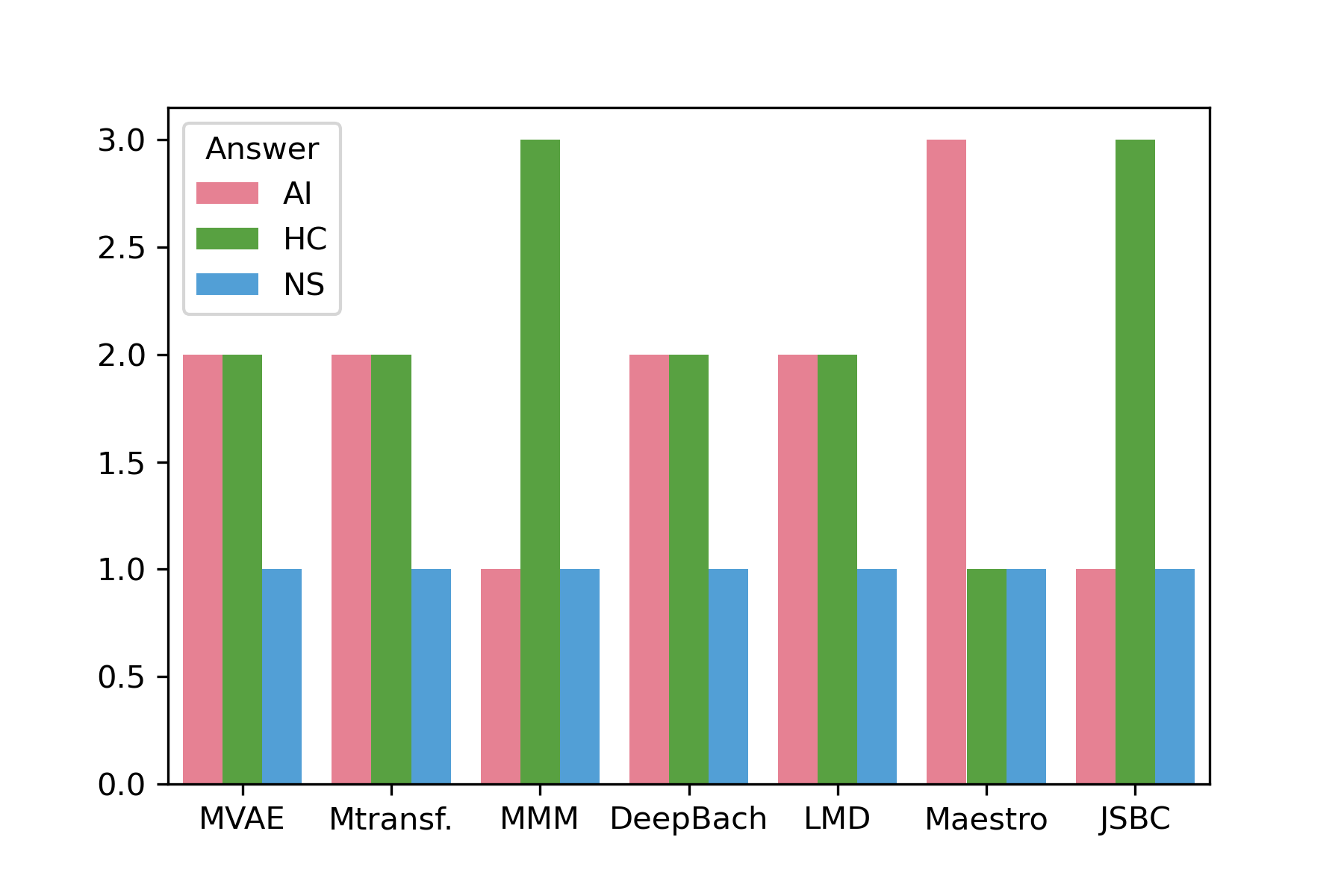}\\
        \caption*{c) Pro user}
    \end{minipage}%
\caption{Turing test for different user levels.}
\label{fig:turing}
\end{figure}

We give the results for 74 beginner, 19 intermediate and 5 pro users from different nationalities. In Fig. \ref{fig:overall} we show the results of the Overall Rating evaluation results and in Fig. \ref{fig:turing} we show the Turing test results per user level. Table \ref{tab:survey_means} shows the means of the numeric answers to the surveys questions by model and dataset. We also show  the results of MuseGAN "from scratch" composer model for the non-pro and pro users \cite{musegan}. Questions Q2, Q3, Q4 and overall rating in our intermediate user survey are the H, R, C and OR questions in MuseGAN user study, respectively.

\begin{table*}[!t]
    \centering
    \caption{Online survey answers per user level and model. The results displayed are the means of the answers. Each question (Q) follows the same order as shown in Table \ref{tab:survey}. Note that R in musegan is the opposite question to our survey question Q3.}
    \begin{tabular}{
    >{\centering\arraybackslash}p{1.2cm}
    >{\centering\arraybackslash}p{3.5cm}|
    >{\centering\arraybackslash}p{0.7cm} 
    >{\centering\arraybackslash}p{0.7cm}|
    >{\centering\arraybackslash}p{0.7cm} 
    >{\centering\arraybackslash}p{0.7cm} 
    >{\centering\arraybackslash}p{0.7cm}| 
    >{\centering\arraybackslash}p{0.7cm} 
    >{\centering\arraybackslash}p{0.7cm} 
    >{\centering\arraybackslash}p{0.7cm} 
    >{\centering\arraybackslash}p{0.7cm} 
    }
    \toprule
    & & \multicolumn{2}{c|}{Beginner} & \multicolumn{3}{c|}{Intermediate} & \multicolumn{4}{c}{Pro}\\
    \midrule
    \multicolumn{2}{c|}{Model or Dataset} & Q1 & Q2 & Q1 & Q2 & Q3 & Q1 & Q2 & Q3 & Q4\\
    \midrule
    \midrule
    \multirow{6}{1cm}{Models} & MusicVAE & 2.99 & 2.86 & 3.00 & 2.84 & 3.61 & 3.60 & 3.60 & 3.20 & 3.40\\
    & MTransf. & 3.51 & 3.35 & 3.95 & 3.22 & 3.17 & \textbf{4.20} & \textbf{4.20} & 2.60 & 3.80\\
    & MMM & 2.99 & 2.85 & 3.42 & 3.47 & 2.63 & \textbf{4.20} & 2.0 & 2.40 & 2.80\\
    & DeepBach & \textbf{3.92} & \textbf{3.70} & \textbf{4.37} & \textbf{4.11} & \textbf{1.79} & 3.00 & \textbf{4.20} & \textbf{1.80} & \textbf{4.20}\\
    & MuseGAN non-pro \cite{musegan} & - & - & - & 3.12 & 1.64 & - & - & - & -\\
    & MuseGAN pro \cite{musegan} & - & - & - & 2.66 & 1.87 & - & - & - & -\\
    \midrule
    \multirow{3}{1cm}{Datasets} & LMD & 3.97 & 3.65 & 3.90 & 4.00 & 2.00 & 4.20 & 4.40 & 1.40 & 4.20\\
    & Maestro & 3.73 & 3.50 & 3.95 & 3.53 & 2.61 & 3.00 & 4.20 & 2.60 & 4.20\\
    & JSB chorales & 3.25 & 2.94 & 3.84 & 3.63 & 2.26 & 4.20 & 3.80 & 1.60 & 4.20\\

         \bottomrule
    \end{tabular}
    \label{tab:survey_means}
\end{table*}

\section{Discussion} \label{sec:discussion}
We showed the survey results by model and dataset in section \ref{sec:eval}. Analyzing the violin plots of the overall rating of the models in Fig. \ref{fig:overall} we can see that there is more deviation in the answers of the beginner users whereas the confidence interval is narrower in the pro-level users plot. This means that pro-level users are more in agreement in their evaluations as we can expect due to their higher music knowledge. The model with the highest mean according to the overall rating question is DeepBach which also obtains a better results than the JSB Chorales dataset which is the training dataset of the model. Beginner and intermediate users also rate DeepBach as the best-performing model in terms of the overall rating. Comparing the results in Table \ref{tab:survey_means} with MuseGAN user study results we can see how the means of the answers are similar even if our surveys have been completed by different users. In terms of the overall rating, MuseGAN results for non-pro and pro users are 3.12 and 2.73 whereas the violin plots of our surveys in Fig. \ref{fig:overall} shows that the best-rated model, DeepBach, has a mean above 4 for the intermediate and pro users. According to the Turing test results shown in Fig. \ref{fig:turing} which is the last question of the survey for all our user levels, pro users select the pieces generated with the MMM model are chosen more times as human composed, whereas beginner and intermediate users select DeepBach model as the closes to the human compositions.
Table \ref{tab:survey_means} shows that all user levels also rate DeepBach as the best-performing model.

\section{Conclusions}
We presented results of some state-of-the-art models for symbolic music generation with specific surveys for different user levels. We designed the surveys to contain pieces generated by the models and pieces taken from the training datasets of the models, so we can also include a Turing test in the surveys. After analyzing the results we can conclude that most specialized models perform better. A clear example of that is DeepBach model which obtained better results overall and it was build to compose chorales in the style of J.S. Bach specifically. Future work should focus on combining objective evaluation with subjective evaluation in order to better compare the outputs of different models and approaches. The combination of subjective and objective methods can led to a standardized music evaluation system.

\section*{Acknowledgments}

This research has been partially supported by the Spanish Science, Innovation and University Ministry by the RTI2018-096986-B-C31 contract and the Aragonese Government by the AffectiveLab-T60-20R project.

\bibliographystyle{ACM-Reference-Format}
\bibliography{sample-base}


\end{document}